\documentclass[
preprintnumbers,
 amsmath,amssymb,
 aps,
 prx,
twocolumn,
superscriptaddress
]{revtex4-2}
\usepackage{lineno}

\usepackage{setspace}
\usepackage{times}
\usepackage{graphicx}
\usepackage{dcolumn}
\usepackage{bm}

\usepackage{xcolor}
\usepackage{upgreek}
\usepackage{comment}
\usepackage{multirow}
\usepackage{color,soul}
\usepackage{caption}
\usepackage{ragged2e}
\usepackage[T1]{fontenc}
   
\makeatletter
\newcommand*{\rom}[1]{\expandafter\@slowromancap\romannumeral #1@}
\makeatother

 \newcommand{\MyTable}
  {
 \centering
  \begin{tabular}{ | c | c | c | c | c | c | }
   \hline
    Sample & Facet & Junction & Metal  & Diameter & Wiring\\ \hline \hline

    S1  & (001) & A & Ti(3 nm)/Au(150 nm) & 340 $\upmu$m & silver paste \\ \cline{3-6}
                         &                        & B & Ti(5 nm)/Au(150 nm) & 340 $\upmu$m & silver paste \\ \hline
   
    S2  & {$\widehat{n} = [0.4, 0.6, 0.7]$} & A & Ti(3 nm)/Au(150 nm) & 340 $\upmu$m & Au wire bonding \\ \cline{3-6}
                         &                        & B & Ti(3 nm)/Au(150 nm) & 340 $\upmu$m & silver paste \\ \hline

  \end{tabular}
  \caption*{\label{Table1}Table 1. Facets, counterelectrodes, diameters, and wiring methods for the junctions in S1 and S2.} 

}

\def\red{\color{black}}
\def\blue{\color{black}}

\begin{document}

\title{Probing $p$-wave superconductivity in UTe$_2$ via point-contact junctions }

\author{Hyeok Yoon}
\author{Yun Suk Eo}
\affiliation{Maryland Quantum Materials Center, Department of Physics, University of Maryland, College Park, MD 20742, USA}

\author{Jihun Park}
\affiliation{Maryland Quantum Materials Center, Department of Physics, University of Maryland, College Park, MD 20742, USA}
\affiliation{Department of Materials Science and Engineering, University of Maryland, College Park, MD 20742, USA}

\author{Jarryd A. Horn}
\author{Ryan G. Dorman}
\author{Shanta R. Saha}
\author{Ian M. Hayes}
\affiliation{Maryland Quantum Materials Center, Department of Physics, University of Maryland, College Park, MD 20742, USA}

\author{Ichiro Takeuchi}
\affiliation{Maryland Quantum Materials Center, Department of Physics, University of Maryland, College Park, MD 20742, USA}
\affiliation{Department of Materials Science and Engineering, University of Maryland, College Park, MD 20742, USA}

\author{Philip M. R. Brydon}
\affiliation{Department of Physics and MacDiarmid Institute for Advanced Materials and Nanotechnology, University of Otago, P.O. Box 56, Dunedin 9054, New Zealand}

\author{Johnpierre Paglione}
\email{paglione@umd.edu}
\affiliation{Maryland Quantum Materials Center, Department of Physics, University of Maryland, College Park, MD 20742, USA}
\affiliation{Canadian Institute for Advanced Research, Toronto, Ontario M5G 1Z8, Canada}


\begin{abstract}
{\red Uranium ditelluride (UTe$_2$) is the strongest contender to date for a $p$-wave superconductor in bulk form.} Here we perform a spectroscopic study of the ambient pressure superconducting phase of UTe$_2$, measuring conductance through point-contact junctions formed by metallic contacts on different crystalline facets down to 250 mK and up to 18~T.
Fitting a range of qualitatively varying spectra with a Blonder-Tinkham-Klapwijk (BTK) model for $p$-wave pairing, we can extract gap amplitude and interface barrier strength for each junction. {\blue We find good agreement with the data for a}  $p_y$-wave gap function with amplitude in 0.26 $\pm$ 0.06 meV. 
{\red Our work provides spectroscopic evidence for a gap structure consistent with the proposed spin-triplet pairing in the superconducting state of UTe$_2$.}
\end{abstract}

\maketitle
 
\section{Introduction}

The recent discovery of spin-triplet superconductivity in UTe$_2$ \cite{Ran:2019} has raised the possibility of realizing the technological dream of odd-parity pairing with non-trivial topology in a natural solid state material. 
The strongest signatures of triplet pairing in UTe$_2$ include upper critical fields greatly exceeding Pauli limits for each crystal orientation \cite{Ran:2019}, re-entrant superconductivity in ultra-high magnetic fields \cite{Ran2:2019},
and near absence of changes in the NMR Knight shift below the superconducting transition temperature  \cite{Ran:2019, Nakamine:2019, Matsumura:2023}.
Together with the observation of chiral in-gap states revealed by scanning tunneling microscopy (STM) studies \cite{Jiao:2020} and a normal surface fluid identified in microwave impedance measurements \cite{Bae2021},
these ingredients provide strong evidence for the non-trivial topological nature of superconductivity in UTe$_2$.

The intrinsic symmetry of the superconducting order parameter, which requires identification in order to understand both the pairing mechanism as well as the nature of topological excitations, belongs to the irreducible representation of the point group of the material's crystal structure ($D_\textrm{2h}$). Assuming spin-triplet pairing, the orbital component of the order parameter should be odd, constraining the possible candidates to $A_u$ (full-gap), $B_{iu}$ ($i$=1,2,3; point-nodes), and their combinations \cite{Shishidou:2021, Ishizuka:2019}. 
Experimental studies on the first generation of UTe$_2$ crystals proposed a 
multi-component order parameter, i.e. $B_{3u} + iB_{2u}$ or $A_{u} + iB_{1u}$, 
based on the observation of nodal excitations in thermal transport \cite{Metz:2019}, broken time-reversal symmetry (TRS) in polar Kerr effect experiments and two distinct superconducting transitions in specific heat \cite{Hayes:2021}. However, later generation materials with higher $T_\textrm{c}$ values appear to have only a single thermodynamic transition, a small but finite Knight shift \cite{Nakamine:2019, Matsumura:2023}, and an apparent lack of TRS breaking \cite{Ajeesh:2023}, raising the possibility that a two-component order parameter is not an intrinsic property of UTe$_2$ \cite{Thomas:2021, Rosa:2022, Sakai:2022}. 
Ultimately, the lack of a jump in elastic shear moduli in both generations of materials \cite{Theuss:2023} points conclusively to a single-component order parameter,
but is still inconsistent with the observation a quadratic temperature dependence of magnetic penetration depth for all crystallographic directions \cite{Metz:2019,Ishihara:2023}.

Spectroscopic studies have historically been very decisive in determining superconducting order parameter symmetry. In UTe$_2$, STM studies by four independent groups have successfully probed the cleaved surface and the superconducting gap at the Fermi level \cite{Jiao:2020,Aishwarya:2023,Aishwarya2:2023,Gu:2023,Lafleur:2023,Hu:2022}, but have only studied the easy-cleavage plane (011) \cite{Jiao:2020, Aishwarya:2023, Gu:2023, Aishwarya2:2023} and the (001) surface \cite{Hu:2022}. More important, studies of the gap structure by STM have been hindered by the abundance of in-gap states that fill in a large fraction of the differential conductance, which remains a mystery but is likely affected by the presence of surface states such as charge- \cite{Aishwarya:2023,Aishwarya2:2023} and pair-density wave orders \cite{Gu:2023}, as well as the anomalous non-superconducting fluid at the surface \cite{Bae2021}.

An alternative approach to studying the directional nature of the superconducting order parameter is to fabricate normal-metal/(insulator)/superconductor (N-(I)-S) junctions in which facets are defined by oriented polishing of bulk single crystals, and performing spectroscopic tunneling experiments.
However, to date the realization of a functional device has been a challenge due to the lack of understanding of surface oxidation and interface quality. In this study, we present the successful measurement of energy spectra in Au/Ti/UTe$_2$ planar junctions formed on two different facet orientations by utilizing the surface oxidation layer of UTe$_2$. The observed conductance spectra are well described by a $p$-wave BTK model for tunneling into triplet superconductors and suggest a $p_y$-wave symmetry as the most plausible order parameter for the ambient pressure superconductivity of UTe$_2$.

 \section{Methods}

\begin{figure} [b]

\includegraphics[width=0.45\textwidth] {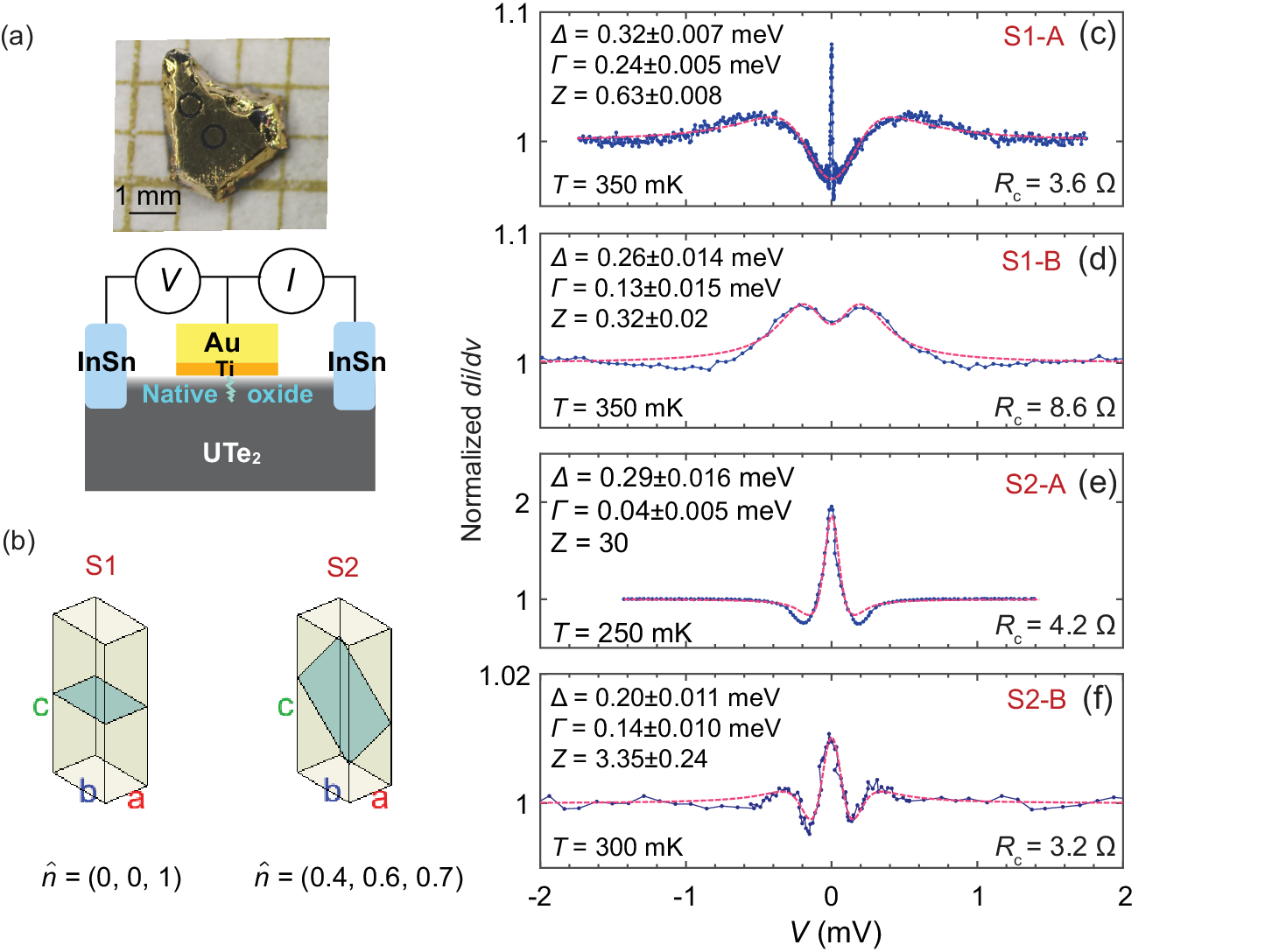} 
    \caption{{\textbf{Differential conductance spectra of UTe$_2$ junctions at 250-350 mK.}} (a) Schematic of point-contact devices fabricated using single-crystal samples of UTe$_2$. (b) Graphical representation of facet directions of samples S1 and S2. (c)-(f) Normalized differential conductance (blue) and fits (magenta) using a $p$-wave BTK model as described in the main text. Obtained fit parameter values for gap energy $\Delta$, scattering rate $\Gamma$, impedance $Z$, and temperature $T$ are noted. Here, $T$ is fixed to the  experimentally measured temperature, and $Z$ is a fitting parameter except for S2-A. {\red Z is set to 30 for S2-A for convenience, since there is very little variation in the spectra with increasing Z upon entering the tunneling regime. Each junction resistance is shown in the figure.}}
\label{fig1} 
\end{figure}

Single crystals of UTe$_2$ were grown by the chemical vapor transport method \cite{Ran:2019,Ran:Synthesis}, yielding samples with a transition temperature $T_\textrm{c}=$ 1.6 K. Orientation of crystal facets was determined using the anisotropy in magnetic susceptibility. Facets on two samples (S1 with (001) facet, and S2 with facet normal vector $\widehat{n} =  [0.4, 0.6, 0.7]$; see SM, section \rom{1}) were polished using aluminum oxide lapping films, and metal contacts were fabricated using Au thin films deposited by evaporation. {\red 3-5 nm of Ti is deposited prior to Au to enhance adhesion to UTe$_2$}. S1 and S2 were  patterned by conventional lithography, with details of the fabrication process and dimensional parameters described in SM, section \rom{2}{\red -A and B}.
During fabrication, samples are heated at 100 $^\circ$C for 1 minute for baking photoresist and additionally heated at 100 - 120 $^\circ$C for 2-3 minutes to adhere glue to a cover glass. 
In-Sn solder was used to make ohmic contacts to samples {\red(see SM, section \rom{2}-C)}. The final device structure is shown in Fig. \ref{fig1}(a). For the electrical measurements, transport and differential conductance measurements were performed using a $^3$He commercial probe with base temperature of 300 mK.

\section{Results}

Our point-contact junctions incorporate the native oxide of UTe$_2$ that forms upon exposure to air. The baking process of photoresist during the fabrication further enhances the surface oxidation, making its thickness more than tens of nanometers {\red(see SM, section \rom{3}).} Despite the thick oxidation layer, our junction resistances maintain low ($R_\textrm{c}$ $<$ 10 $\Omega$) values at low temperatures. Considering the nature of the oxidation layer, it is likely that there exists metallic shorts through the layer that form  leaking paths in large-size N-I-S planar junctions that decrease the effective size of junctions, thereby suppressing inelastic electron scattering across the junctions (Fig. \ref{fig1}(a)). In this ballistic regime, the spectra can reflect the energy density of states of the sample layer as shown in many other cases \cite{Yanson:1974, Naidyuk:2003, Hwang:2015, Srikanth:1992, Dvoranova:2018, Hagiwara:1973}.
Despite the inability to form a proper tunnel barrier, we can therefore utilize this configuration to perform point-contact spectroscopy.

\begin{figure} [b]
\includegraphics[width= 0.45\textwidth]{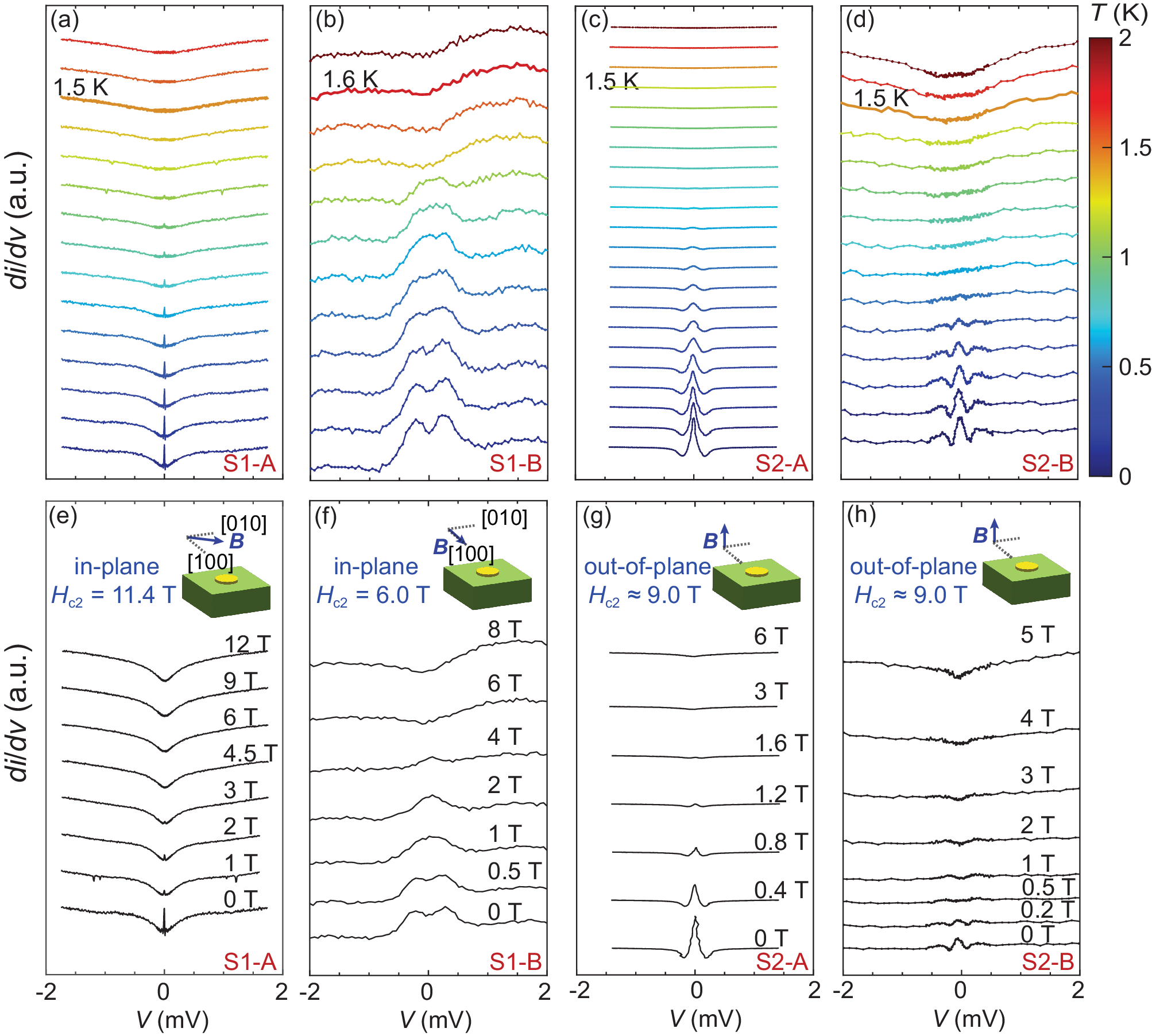}
\caption{\textbf{Temperature and magnetic field dependence of the differential conductance of UTe$_2$ junctions} (a)-(d) Zero-field differential conductance $di/dv$ measured at fixed temperatures between 0.3~K and 2 K. For each panel, the $T=$ 1.5 K or 1.6 K data are labeled to denote the resistive transition temperature. (e)-(h) Magnetic field dependence of $di/dv$ measured at 350~mK for S1 and 250~mK for S2. For S1-A and S1-B, the magnetic field is applied in-plane, and the upper critical field is obtained from the independently measured resistive transition. For S2-A and S2-B, the magnetic field is applied out-of-plane, and the upper critical field is estimated using the orientation of facets and the known angle dependence (see SM, section \rom{5}). }
\label{fig2}
\end{figure}

Figure \ref{fig1} and \ref{fig2} show the differential conductance spectra across the junctions (S1-A, S1-B, S2-A, S2-B) at the base temperature and their temperature evolution. Here, S1 and S2 label the crystals, and A and B label the junctions. For example, S1-A is the junction A fabricated on the crystal S1. S1-A and S1-B are fabricated on the same facet (001) of the crystal S1, and S2-A and S2-B are created on the opposing faces of crystal S2, which run parallel to each other. Figure \ref{fig1}(b) presents a schematic of the facets of the junctions on samples S1 and S2, demonstrating the different orientation of the conductance measurements for each sample. In Fig. \ref{fig1}(c)-(f), spectra are normalized to the normal-state spectra above $T_\textrm{c}$. As shown in Fig. \ref{fig2}, features in the differential conductance spectra are developed below the superconducting transition temperature and the upper critical field for all junctions, and the details of each spectra are described below. 

All junctions presented in this study exhibit spectral features consistent with a gap opening of UTe$_2$. Sample S1-A exhibits a dip feature at an energy close to the expected superconducting gap energy $\Delta = 0.25$ meV, estimated from the weak-coupling BCS theory ({\it i.e.}, $2\Delta/k_\textrm{B}T_\textrm{c}$=3.56), with shoulders reminiscent of coherence peaks in the superconducting density of states. However, this sample also exhibits a prominent peak at zero-bias, as shown in Fig. \ref{fig1}(c). The coherence peaks vanish in the vicinity of $T_\textrm{c}$, and the parabolic background remains in the normal state spectra. On the other hand, when the magnetic field is applied in-plane (15 $^\circ$ off from the $b$-axis), the coherence peak disappears. We note that determining when exactly the coherence peaks disappear is not clear since the background dip structure is deeper and remains even beyond $\mu_0 H_{c2} = 11.4$ T, in contrast to the normal-state spectra above $T_\textrm{c}$.   
%
While the zero-bias peak (ZBP) is a very interesting feature that could possibly be associated with zero-energy Andreev states which occur 
in a topological superconductor, we first note that its energy width is narrower 
than the minimum broadening possible due to the thermal smearing of the spectra; the peak width is $\sim 10$ $\upmu$eV while $k_\textrm{B} T$ = 25 $\upmu$eV at $T =$ 300 mK. Out of many possible reasons for the ZBP in the differential conductance \cite{Kuerten:2017}, to our knowledge the only source immune to thermal broadening is from a Josephson supercurrent across the junction. 
This is an interesting aspect of sample S1-A that was not reproduced in other junctions, and may be an indication of a tunneling phenomenon. However, it is important to note that {\red our Au/Ti counter-electrode is not superconducting in the range where the zero-bias peak is observed (i.e. up to 1 K), } raising the question of what component plays the role of a superconducting electrode in such a SNS or SIS Josephson configuration \footnote{For our junctions, whereas the ZBP ceases to exist at around $T_\textrm{c}$, it disappears earlier than $\mu_\textrm{0}H_{c2}$ at around 3.5 T in response to the in-plane magnetic field. The maximum value of the ZBP has an oscillation with respect to the in-plane magnetic field, which resembles the Fraunhofer pattern shown in the Josephson junction (SM, section \rom{4}). This observation supports the Josephson supercurrent hypothesis as the origin of the ZBP. The estimated junction area is 0.008 $\upmu$m$^2$ when the first lobe approximately ends at 0.25 T. Interestingly, it has been reported that the Josephson effect is observed with normal-metal electrodes in UBe$_{13}$ point-contact. The origin is discussed as a proximitized normal layer or phase-slip \cite{Wolf:1987, Kadin:1990}, and our observation may share the origin with that.}.

Sample S1-B, which is a separate junction on the same crystal facet as S1-A, exhibits a different shape consisting of a small dip imposed on a broad conductance enhancement (Fig. \ref{fig1}(d)). In addition, we also see a weak dip in conductance at energies higher than the low-energy enhancement. This combined peak-dip structure is often observed in the thermal regime of contacts and can be explained by the influence of the critical current, causing the system to transition into a high-resistance state before the local superconductivity is suppressed \cite{Kumar:2021,Sheet:2004, Aggarwal:2016, Aggarwal:2019}. In this case, the small dip represents the energy spectra imposed on the broad peak due to the superconductivity. The energy scale of this dip feature matches the superconducting gap size $\Delta$ ($\sim0.2$ meV) of UTe$_2$, which is consistent with this scenario.  Note that the peak-dip feature has also been observed in other materials such as Pt-Sr$_2$RuO$_4$ point-contact junctions, but was explained in a different manner by incorporating a phenomenological transmission cone instead of attributing the structure to critical current effects \cite{Laube:2000}. This model suggests that it could be also feasible to fit our S1-B junction without taking the critical current into account. 

Sample S2-A exhibits a peak-dip structure as well, but with a very sharp ZBP and dips on either side at higher energies. The strong ZBP has amplitude nearly twice as large as the background as shown in Fig. \ref{fig1}(e), and is often seen in point-contact junctions and explained via various origins. First, the ZBP may result from the prevalent Andreev reflection. However, the distinctive dip outside this ZBP can not be explained solely with Andreev reflection. {\red  It is possible that the peak-dip structure is originated from the addition of tunneling and Andreev reflection in the presence of proximitized normal layer \cite{Strijkers:2001}. However, we rule out this case because the energy scale of the peak is close to the UTe$_2$ gap, unlike the expectation from this scenario. It  is also possible that the ZBP can appear in s-wave/ferromagnet interface, but we exclude this scenario because our junctions do not consist of such ingredients   \cite{Usman:2011}.} Second, the peak-and-dip structure can be interpreted as the effect of critical current, as discussed in the section for S1-B. Nonetheless, we exclude this scenario because the background of the spectra is {\red $T$ independent when the ZBP appear, whereas the transport-like conduction is expected to show the resistivity decreasing quadratically in temperature (see SM, section \rom{6}).} In addition, if the dip and ZBP structure arises from the critical current, the dip should be spike-like and its position should shift to the center as {\red $T$ increases}, as reflecting the features of the critical current. In contrast, the width of the ZBP in our spectra does not change significantly with increasing {\red $T$}. Also, in our case, the peak intensity is sharply suppressed as {\red $T$} increases, contrasting with the binary character of the resistive transition.  Hence, we attribute the ZBP to the existence of surface Andreev-bound states often observed in the tunneling limit. This ZBP originates from the interference of the transmitted electron-like quasi-particle and hole-like quasi-particle experiencing the phase difference of the pair potential~\cite{Hu:1994, Tanaka:1995}. This constructive interference can be induced when the electrons are injected along the nodal direction of $d$-wave superconductors \cite{Daghero:2012} or in the $p$-wave topological superconductors~\cite{Sakai:2011, Zhu:2023}.

Finally, for sample S2-B, the spectrum has similar features to both S1-B and S2-A, but with much smaller ZBP intensity as compared to S2-A as shown in Fig. \ref{fig1}(f). 
The varying features in the four spectra present differing behavior that are actually useful for identifying distinct responses due to gap structure, and can be reasonably well modeled by variations of junction parameters using a $p$-wave gap scenario as explained below. 

\begin{figure} [t]
\includegraphics[width=0.45\textwidth]{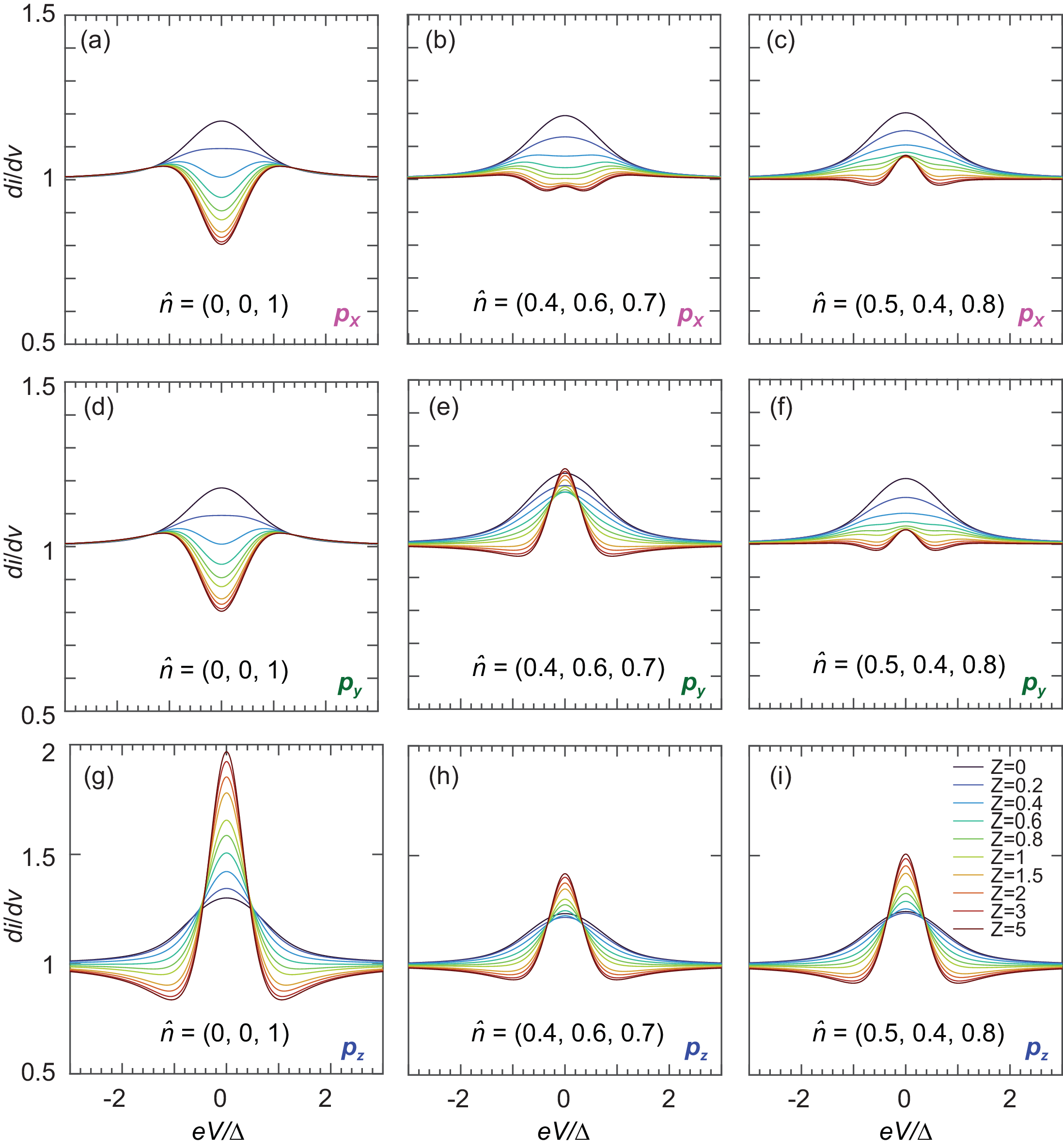}
\caption{\textbf{BTK simulation of normalized differential conductance using $p$-wave model.} Simulated conductance spectra are presented for junctions oriented along different facets of the orthorhombic crystal structure, plotted as a function of interface barrier strength $Z$ (color scale) ranging between 0 (blue) and 5 (red). (a)-(c), (d)-(f), and (g)-(i) show the BTK simulation with $p_x$, $p_y$, $p_z$,  respectively. (a), (d), (g) are BTK simulation at the facet $\widehat{n} =  [0, 0, 1]$.  (b), (e), (h) are BTK simulation at the facet $\widehat{n} =  [0.4, 0.6, 0.7]$. (c), (f), (i) are BTK simulations at the facet $\widehat{n} =  [0.5, 0.4, 0.8]$. All spectra are plotted with fixed parameters $k_BT=0.1\Delta$ and $\Gamma=0.4\Delta$. }
\label{fig3}
\end{figure}

To model these conductance spectra, we have utilized a generalization of the BTK theory~\cite{BTK:1982} for tunneling into triplet superconductors~\cite{Yamashiro1997}, calculating the conductance according to the formula  
\begin{equation}
\frac{dI}{dV} = \sigma_N\int^{\infty}_{-\infty}\sigma_{\text{BTK}}(E+i\Gamma)\left[-\frac{\partial f(E+eV)}{\partial E}\right]dE
\end{equation}
where $\sigma_N$ is the normal-state tunneling conductance, $\sigma_{\text{BTK}}(E)$ is the normalized BTK conductance, $f(E)$ is the Fermi function, and $\Gamma$ is a phenomenological broadening parameter. Details of the calculation are presented in the supplemental material (see SM, section \rom{7}). {\red As discussed earilier, a number of previous experiments point to a triplet order parameter in UTe$_2$, and so we assume a $p$-wave {\blue triplet} model to fit the data. We also find that $p$-wave symmetry gives a better fit than $s$-wave and $d$-wave {\blue singlet states,} as the $p$-wave fit results in the smallest $\Gamma/\Delta$ value. (see SM, section \rom{8}).} The orthorhombic crystal symmetry of UTe$_2$ places few constraints on the structure of the triplet pairing states in each irrep: even upon restricting to $p$-wave gap symmetry, a general pairing state has two ($B_{iu}$, $i=1,2,3$) or three ($A_{u}$) independent gap components. {\red Since our experimental results are unable to distinguish between single-component (non-chiral) and multi-component (chiral) gaps due to the limited energy resolution,  we assume that the pairing state preserves TRS to keep our task manageable}.
Our fitting parameters hence consist of the relative strength of the different $p$-wave components, the overall gap amplitude $\Delta$, the interface barrier strength $Z$, and  $\Gamma$. {\blue We find that the data is best fit by a pairing state with a dominant $p_y$-wave gap component.} {\red In other words, our spectra is best fitted with $B_{1u}$, $B_{3u}$, and $A_{u}$, which all include a $p_y$ component.}
{\blue We are not able to further refine the gap structure because of the low energy resolution resulting from the thermal and intrinsic broadening. Due to this uncertainty, here we present fits to the data} with a purely $p_y$-wave gap function $\Delta({\bf k}) = \Delta_0 k_y/k_F$ with zero-temperature gap amplitude in the range $0.2$ to $0.35$~meV. 
{\red The obtained gap size} 
is consistent with values of $\Delta =  \pm 0.25$~meV obtained by spectroscopic STM experiments on the [011] crystalline surface \cite{Jiao:2020}. We stress that we cannot exclude the presence of other symmetry-allowed $p$-wave gap components, {\blue which are required for consistency with thermodynamic measurements indicating point nodes, but} our analysis suggests that they are {\blue smaller than the $p_y$-wave component.}

The ZBP offers an important clue to gap structure of an unconventional superconductor: a pronounced ZBP typically indicates
zero-energy surface Andreev bound states, which occur when the superconducting gap changes sign upon reversing the momentum component normal to the surface; on the other hand, the absence of the ZBP is consistent with a gap which a does not change sign upon this reversal. For a $p$-wave superconductor the differential conductance is anisotropic; in particular, the presence or absence of a ZBP at a given surface is characteristic of different $p$-wave gaps. {\blue This argument is approximately robust to the presence of subdominant gap components, as discussed in the supplemental material (see SM, section \rom{7}).}
To visualize this concept, Fig. \ref{fig3} demonstrates the normalized differential conductance spectra for the facets of our samples, simulated from the BTK theory using a $p$-wave model. As shown in Fig. \ref{fig3} (a), (d), and (g), the gap-like feature around zero-bias in the S1-A and S1-B conductance data ($\widehat{n} =  [0, 0, 1]$) is thus not consistent with a dominant $p_z$-wave gap component, and indeed can be reasonably fit by a purely $p_x$- or $p_y$-wave gap. 

On the other hand, as shown in Fig. \ref{fig3} (b), (e), and (h), a purely $p_z$- or $p_y$-wave state gives the best fit to the sharp ZBP in the S2-A data, consistent with the surface normal lying $23^\circ$ away from the $y$-$z$ plane. We note that the prominence and height of the peak is enhanced by reducing the broadening $\Gamma$, which in Fig.~\ref{fig3} is several times larger than in our fit to the S2-A data. Since the S1 data excludes a dominant $p_z$-wave component, this implies that the gap is predominantly $p_y$-wave in character. 

The S2-B case is least consistent with a purely $p_y$-wave gap, and a better fit is obtained for a $p_x$-wave state.
Although this is hard to reconcile with the S2-A data for nominally the same surface normal, assuming a slight misalignment of the surface normal compared to the S2-A surface yields an excellent fit to a purely $p_y$-wave gap. Specifically, best agreement is obtained for an approximately $10^\circ$ misalignment ($\widehat{n} =  [0.5, 0.4, 0.8]$), which is shown in Fig. \ref{fig3} (c), (f) and (i). This degree of misalignment is experimentally possible since S2-B is fabricated on the other side of S2-A, possibly making the two facets not exactly parallel. 

\begin{figure}
\includegraphics[width=0.45\textwidth]{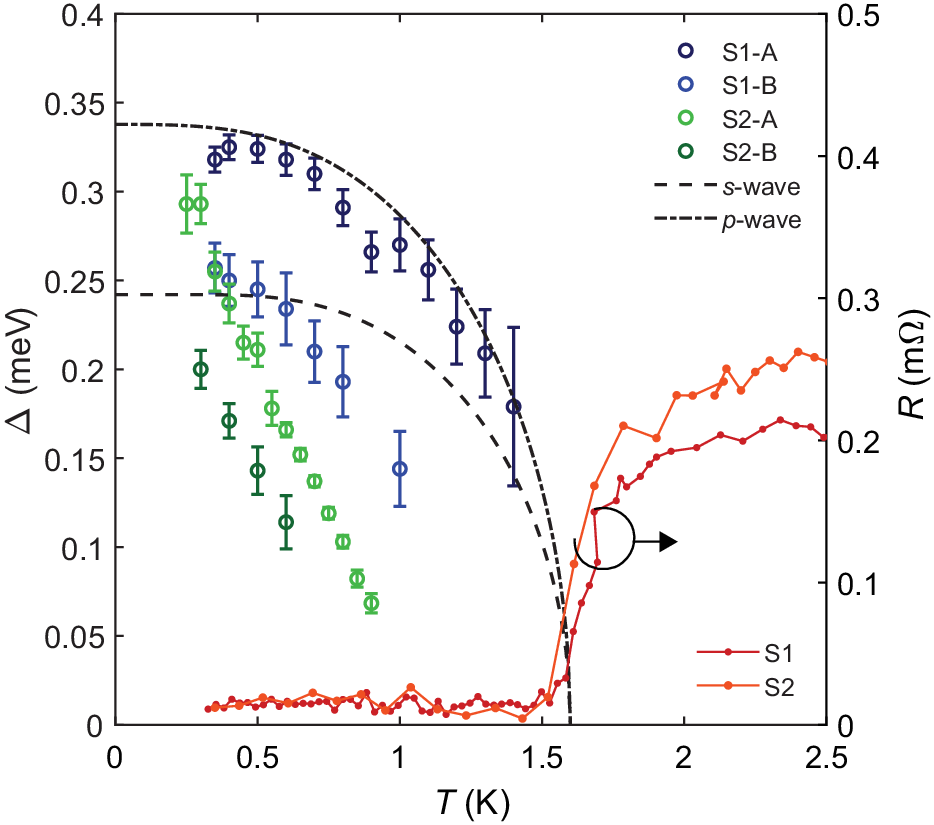}
\caption{\textbf{Extracted Superconducting energy gap magnitudes.} Circles represent the gap sizes extracted from the $p$-wave BTK fits. The uncertainty is defined by the 95$\%$ confidence interval for the fit. The dashed and dotted curves are the simulated temperature dependence of $\Delta$ for $s$-wave and $p$-wave models, respectively, with $T_\textrm{c}$ fixed at 1.6~K.  The red and orange curves, measured in S1 and S2, respectively, demonstrate the superconducting transition measured by four-probe resistance using ohmic contacts as current leads and point-contacts for voltage leads. }
\label{fig4}
\end{figure}

The variation of the fitted gap amplitude with temperature is shown in Fig.~\ref{fig4}. The S1-A data follows rather closely the weak-coupling temperature dependence of a $p$-wave pairing state with $T_c=1.6$ K; the S1-B data shows a similar variation albeit with somewhat lower critical temperature. In contrast, the S2-A and S2-B data show an approximately linear decline in the gap as a function of temperature, extrapolating to zero for $T_c\leq1$ K, which suggests a lower $T_c$ at the surface. 

Our theoretical analysis has utilized a number of standard simplifying assumptions. In particular, we treat the Fermi surface in both the lead and superconductor as spherical with the same Fermi wavevector and effective mass. 
Although this is inconsistent with the quasi-2D Fermi surface and significant orthorhombic anisotropy in normal-state resistivity in UTe$_2$, accurately accounting for the band structure typically does not introduce significant quantitative changes in the BTK theory predictions~\cite{Daghero_2010}. However, the unexpected linear $T$-dependence of the gaps measured at the S2 surfaces may indicate a breakdown of our BTK theory. In particular, we have neglected the variation of the gap near the surface of the material, which could be significant at the S2 surfaces of our proposed $p_y$-wave state. 
Accounting for this might alter the quantitative values of our fit parameters but is not expected to qualitatively alter our conclusions; in particular, the relation between the ZBP and the gap symmetry can be formulated in terms of topological invariants~\cite{Schnyder_2015}, making this feature somewhat immune to details of the surface.

In summary, we have presented point-contact spectroscopy spectra of the superconducting state of UTe$_2$ using four distinct junctions fabricated by depositing Ti/Au metal contacts on the native oxide surface of UTe$_2$ single crystals with two different facet orientations. By fitting conductance spectra measured with currents directed along both ($\widehat{n} =  [0, 0, 1]$) and  ($\widehat{n} =  [0.4, 0.6, 0.7]$), we are able to model the data with a simple $p$-wave BTK model, extracting gap amplitude and constraining the gap structure.
All junctions exhibit spectroscopic features that close at the superconducting transition temperature and upper critical field of UTe$_2$, with energies in 0.26 $\pm$ 0.06 meV consistent with energy scales observed in scanning tunneling spectroscopy and derived from thermodynamic quantities.
Upon careful examination of a $p$-wave BTK model, we conclude that a gap with a dominant $p_y$-wave component is the most consistent with our data. 
Our study demonstrates the potential of performing electronic spectroscopy in UTe$_2$ in a stable device with choice of crystalline facet direction and external environment, opening the door to further studies of the multiple superconducting phases of UTe$_2$, including the re-entrant and field-polarized states. 

\section*{Author contributions}
HY and JP (J.Paglione) conceived and designed the experiments. 
SRS and IMH synthesized single crystals. 
HY and RGD characterized the orientations of single crystals. 
HY and YSE fabricated samples to make point-contact junctions. 
HY, YSE, and JAH performed the electrical measurements. 
HY and JP (J.Park) characterized the surface oxidation of the samples. 
PMRB provides the theoretical modeling and fitting of data.
HY, IT, PMRB, JP (J.Paglione) wrote the paper. All authors read and approved the final manuscript.

\section*{Acknowledgements}
We acknowledge useful discussions with D. F. Agterberg, V. Madhavan, {\red and A. R. Hight Walker.} Research at the University of Maryland was supported by the U.S. Department of Energy Award No. DE-SC-0019154 (sample characterization), 
the Air Force Office of Scientific Research under Grant No. FA9950-22-1-0023 (spectroscopic experiments), 
the Gordon and Betty Moore Foundation’s EPiQS Initiative through Grant No. GBMF9071 (materials synthesis), and the Maryland Quantum Materials Center. S.R.S. acknowledges support from the National Institute of Standards and Technology Cooperative Agreement 70NANB17H301.
P.M.R.B. was supported by
the Marsden Fund Council from Government funding, managed by Royal Society Te Ap\={a}rangi, Contract No. UOO1836.

\section*{Competing interests}
All authors declare no financial or non-financial competing interests. 

\section*{Data availability}
The data that support the findings of this study are available from the corresponding author upon reasonable request.

\newpage


\clearpage
\bibliography{ms}
\bibliographystyle{Science}

\onecolumngrid
\clearpage


\begin{center}
\fontsize{18}{12}\selectfont Supplementary Materials for \break 
\fontsize{14}{12}\selectfont Probing $p$-wave superconductivity in UTe$_2$ via point-contact junctions \break 

\fontsize{12}{12}\selectfont Hyeok Yoon, Yun Suk Eo, Jihun Park, Jarryd A. Horn, Ryan G. Dorman, Shanta R. Saha, Ian M. Hayes, Ichiro Takeuchi, Philip M. R. Brydon, Johnpierre Paglione \break 

{\fontsize{12}{12}\selectfont Correspondence to: paglione@umd.edu }

\end{center}
This PDF file includes: \newline
Figs. S1 - S12. \newline
Captions for Figs. S1 - S12. \newline
Table S1, S2.\newline
Captions for Table S1, S2. \newline
\clearpage

\section*{\rom{1}. The facets of samples} \label{Appendix:Facets}
\begin{spacing}{1.5}
 We use the anisotropy in the magnetic susceptibility $\chi$ to characterize the orientations of the UTe$_2$ crystals. For S1, the facets include the high-symmetry axis, as shown in Fig. S1. The facet of the junctions in S1 is parallel to $a$ and $b$-axis and perpendicular to $c$-axis. On the other hand, for S2, the facet is not parallel to any of the high-symmetry axes. Therefore, we develop the way to find the facet as described below.

 \begin{figure}[h!]
\includegraphics[width= 85 mm]{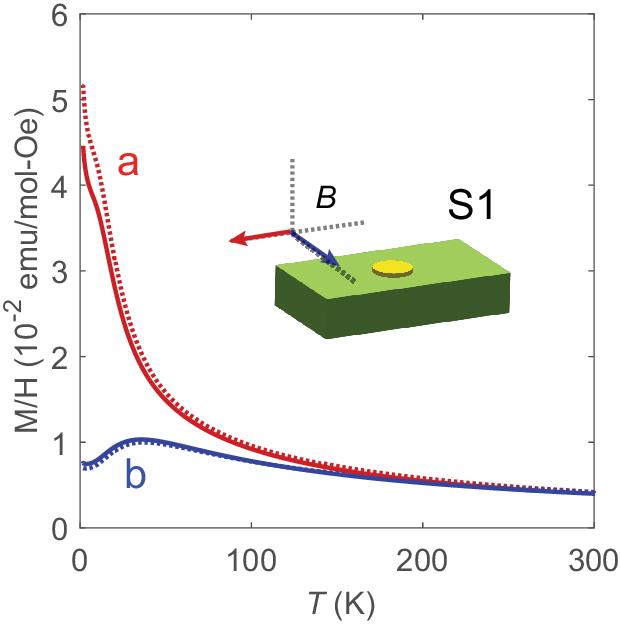}
\caption*{FIG. S1. The determination of the crystal orientation for S1. Two in-plane magnetic susceptibilities measured in S1 matched with $\chi_{aa}$ and $\chi_{bb}$. Each $B$ field direction is represented in the inset. Solid lines are our data in S1, and dotted lines are $\chi_{aa}$ and $\chi_{bb}$ adopted from S. Ran, et al. \cite{Ran:2019}.}
\label{figS1}
\end{figure}

 Consider the initial principal axes of coordinates $x$,$y$,$z$ corresponds to the $a$,$b$,$c$ axis of UTe$_2$ crystal structure. Then, the magnetic susceptibility tensor can be represented as the following. 

 \begin{equation}
\chi = 
\begin{pmatrix}
\chi_{aa} & 0 & 0 \\
0 & \chi_{bb} & 0 \\
0 & 0 & \chi_{cc}
\end{pmatrix}
 \end{equation}

where $\chi_{aa}$, $\chi_{bb}$, $\chi_{cc}$ are the magnetic susceptibilities  in $a$, $b$, $c$ directions when the magnetic field $B$ is applied in $a$, $b$, $c$ directions, respectively. Their temperature dependence is reported in S. Ran et al. \cite{Ran:2019}.
 When the axes are rotated around $z$-axis by $\theta$, the magnetic susceptibility in the new axes $\chi^{'}$ can be represented as the following change of basis.
  \begin{equation} \label{eq2}
\chi^{'} =  \mathbf{R}_{z,\theta}\chi\mathbf{R}_{z,\theta}^\mathrm{T}
 \end{equation}
 where the rotation matrix $\mathbf{R}_{z,\theta}$
\begin{equation*}
\mathbf{R}_{z,\theta} = 
\begin{pmatrix}
\cos \theta & -\sin \theta & 0 \\
\sin \theta & \cos \theta & 0 \\
 0 & 0 & 1
\end{pmatrix}
\end{equation*}

\begin{figure}[t!]
\includegraphics[width= 85 mm]{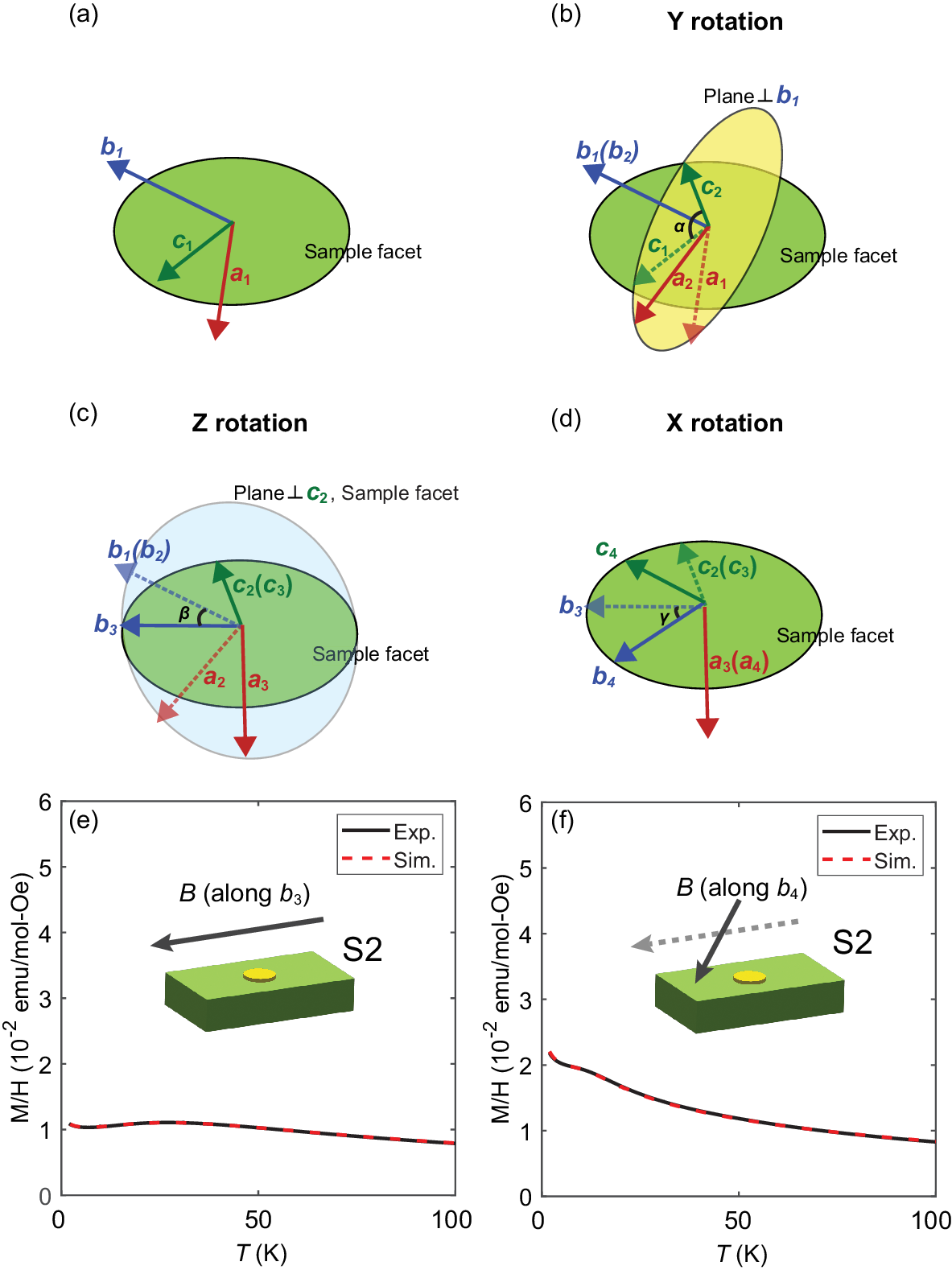}
\caption*{FIG. S2. The determination of the crystal orientation for S2 using $YZX$ Euler rotations. (a) $a_1$, $b_1$, $c_1$ are the crystallographic $a$, $b$, $c$ axes in UTe$_2$. The green plane represents the sample facets. (b) The new coordinate system, denoted by $a_2$, $b_2$, and $c_2$, is obtained by transforming the original coordinates $a_1$, $b_1$, and $c_1$ through a rotation around the $Y$-axis by the angle $\alpha$ -- in this case, around the $b_1$-axis. Here, $\alpha$ is is set to ensure that $c_2$ lies in the sample facet. (c) The new coordinate system, denoted by $a_3$, $b_3$, and $c_3$, is obtained by transforming the original coordinates $a_2$, $b_2$, and $c_2$ through a rotation around the $Z$-axis by the angle $\beta$. Here, $\beta$ is is set to ensure that $b_3$ lies in the sample facet. Also, $a_3$ becomes perpendicular to the sample facet as a result. The light blue plane represents the plane perpendicular to both $c_2$ and the sample facet. (d) The new coordinate system, denoted by $a_4$, $b_4$, and $c_4$, is obtained by transforming the original coordinates $a_3$, $b_3$, and $c_3$ through a rotation around the $X$-axis by the angle $\gamma$. (e) shows the experimental data of the magnetic susceptibility and the fit of the data for the most $b$ axis-like magnetic susceptibility in response to the in-plane $B$ field. In $YZX$-rotation model, this corresponds to the case when $B$ field is applied along $b_3$ axis described in (c) -- In other words, $\gamma$ = 0. The fit parameters are $\beta$ = 17$^\circ$  and $\gamma$ = 0$^\circ$. Any $\alpha$ gives the same fit. (f) shows the experimental data of the magnetic susceptibility and the fit of the data when $B$ field is applied 41 $^\circ$ off from the measurement in (e). In $YZX$-rotation model, this corresponds to the case when $B$ field is applied along $b_4$ axis described in (d). The fit parameters are $\alpha$ = 67$^\circ$, $\beta$ = 16$^\circ$, $\gamma$ = 33$^\circ$. }
\label{figS2}
\end{figure}

This representation in the new coordinate can be generalized to three rotations around $y$, $z$, and $x$ axes with the Euler's angles $\alpha$, $\beta$, and $\gamma$.  

  \begin{equation} 
  \label{eq3}
\chi^{'} =  \mathbf{R}_{x,\gamma} \mathbf{R}_{z,\beta}\mathbf{R}_{y,\alpha}\chi\mathbf{R}_{y,\alpha}^\mathrm{T}\mathbf{R}_{z,\beta}^\mathrm{T}\mathbf{R}_{x,\gamma}^\mathrm{T}
 \end{equation}

In UTe$_2$, the magnetic susceptibility along $b$-axis, $\chi_{bb}$ has a distinctive downturn below $\sim$ 20 K whereas $\chi_{aa}$ and $\chi_{cc}$ increase monotonically with decreasing temperature. Based on this fact, we first find the most $b$ axis-like curve when $B$-field is applied in-plane, as shown in Fig. S2(e). This corresponds to $b_3$ in Fig. S2(c). Then, $b_3$ becomes the projection of the crystallographic $b$-axis ($b_1$ in Fig. S2(a) or equally $b_2$ in Fig. S2(b)) onto the sample facet. By fitting the magnetic susceptibility using eq. (\ref{eq2}), we can find the projection angle $\beta = 17^{\circ}$ which nicely fit the experimental data as shown in Fig. S2(e). With this data, crystallographic $a$ and $c$ are not constrained. 

Next, in order to find $a$ and $c$ axis, we need to fit another curve measured with in-plane $B$ field shown in Fig. S2(d). This is $41^{\circ}$ degree off from the most $b$-like axis, and this is expected $\gamma$. Fitting the corresponding curve using eq. (\ref{eq3}) results in ($\alpha$, $\beta$, $\gamma$) = ($67^{\circ}$, $16^{\circ}$, $33^{\circ}$). $\beta$ is nicely matched with the one obtained from Fig. S2(e), indicating that this method of finding randomly oriented facets is reliable. $\gamma$ is off by 8 $^{\circ}$ from the expected angle measured under the microscope. This difference can come from a small misalignment in sample mounting. These fit parameters can be converted to the normal vector of the facet $\widehat{n} = [0.4, 0.6, 0.7]$.

\clearpage
\section*{\rom{2}. Sample Information} \label{Appendix:SampleInfo}
\subsection{Structural parameters for junctions}

\begin{table*}[h]
\MyTable
\end{table*}

{\red
\subsection*{B. Surface topography of UTe$_2$ single crystals after polishing}
Figure S3. shows the surface topography of referential UTe$_2$ samples that we polish with 0.3 $\upmu$m of aluminium oxide polishing pad, which is the same method for junction fabrications in this study. In the window of 30 $\upmu$m by 30$\upmu$m, RMS roughness is 3.4 nm.
\begin{figure}[h!]
\includegraphics[width= 85 mm]{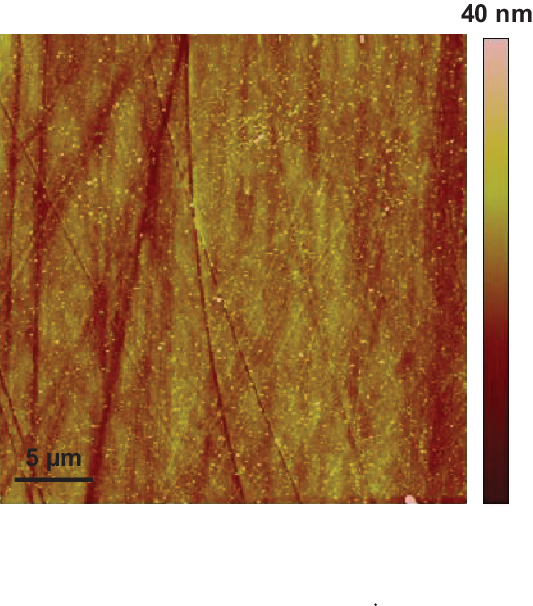}
\caption*{{\red Fig. S3. Surface topography of polisehd surface of UTe$_2$ measured by atomic force microscopy.}}
\label{figS1}
\end{figure}
}
{\red 
\subsection*{C. InSn solder ohmic contact }
 Figure S4. shows the differential conductance taken between two InSn contacts on UTe$_2$ single crystals at 250 mK. $I_+$($I_-$) and $V_+$($V_-$) are shorted in the sample stage, so that the measured resistance consists of UTe$_2$ resistance and two independent contact resistances. The resistance stays around 86 m$\Omega$ without showing the any spectroscopic feature. Also, this value is a few orders of magnitude smaller than the junction resistance including the Au/Ti/UTe$_2$ point-contact. Therefore, we can safely disregard the effects between InSn-UTe$_2$ contact on the spectroscopic features in our spectra}

\begin{figure}[h!]
\includegraphics[width= 85 mm]{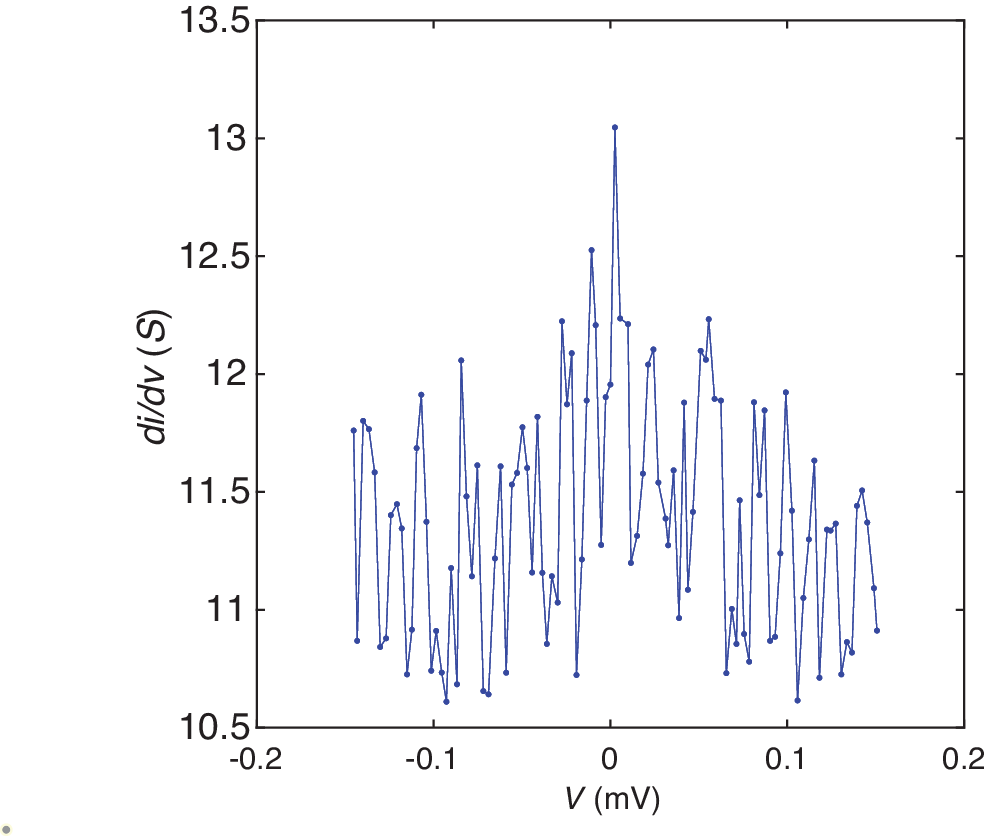}
\caption*{{\red FIG. S4. The differential conductance measurement between two InSn contacts in UTe$_2$ single crystals. }}
\label{figS1}
\end{figure}
\

\clearpage
\section*{\rom{3}. Surface oxidation in UTe2}
{\red
 We use Raman spectroscopy and spectroscopic ellipsometry to understand the interface of UTe$_2$ junctions in this study. First, Raman spectroscopy shows that the heat treatment during the sample fabrication process induces more profound oxidation than the ambient conditions. To measure the Raman peaks in the intrinsic UTe$_2$, the bulk sample was cleaved right before the measurement. The intrinsic UTe$_2$ has a Raman-active mode around 155 cm$^{-1}$, as shown in Fig. S5(a). After 48 hrs in air, the new peaks at 124 and 142 cm$^{-1}$ appear in addition to the original peaks, as shown in Fig. S5(b). These modes can be either $\alpha$-TeO$_2$ or elemental Te, both of which are the semiconductors \cite{Keerthana:2024}.  On the other hand, when the cleaved sample is exposed to heat at 100 $^\circ$C for 5 minutes in air, which is the similar conditions to the sample fabrication process, the oxidation peak becomes more pronounced.
 
 Furthermore, we estimate the thickness of the oxide layer of our sample using spectroscopic ellipsometry. Figure S6 shows the inverse tangential component of the amplitude ratio $\Phi$ at the left-axis and the phase shift $\delta$ at the right-axis. Assuming the oxide layer added on the bare surface of UTe$_2$ after heat-treatment is uniform, we use a bi-layer model which uses a bottom layer as a bare UTe$_2$ spectra (blue) to fit the heat-treated data (red). The best-fits occurs when the thickness of oxide layer is 85 nm.

 For Raman spectroscopy, LabRam Aramis from Horiba Jobi Yvon was used with 532~nm source wavelength, 11$\mu$W power and 200 second exposure time for the measurement. We confirmed that this beam condition does not damage the samples by checking the reproducibility in the spectra in identical conditions. For ellipsometry, M-2000 spectroscopic ellipsometer from J.A. Wollam Co. was used for the measurement, and CompleteEASE software was used for fitting.
 }
 
  \begin{figure}[h!]
\includegraphics[width= 85 mm]{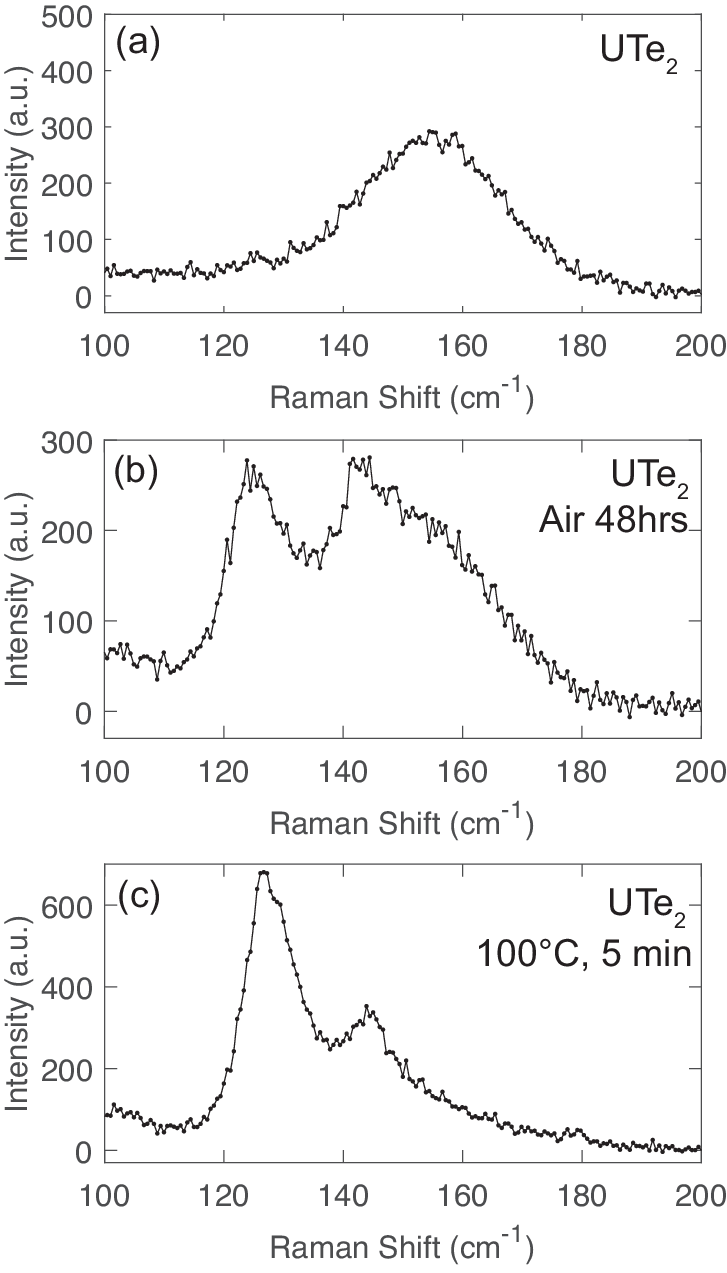}
\caption*{{\red FIG. S5. Raman spectra in (a) intrinsic (b) air-exposed, and (c) heat-treated UTe$_2$. For (b), the exposure time is 48hrs, and for (c), the exposure time is 5 min at 100 $^\circ$C in air.}}
\label{figS4}
\end{figure}

\begin{figure}[h!]
\includegraphics[width= 85 mm]{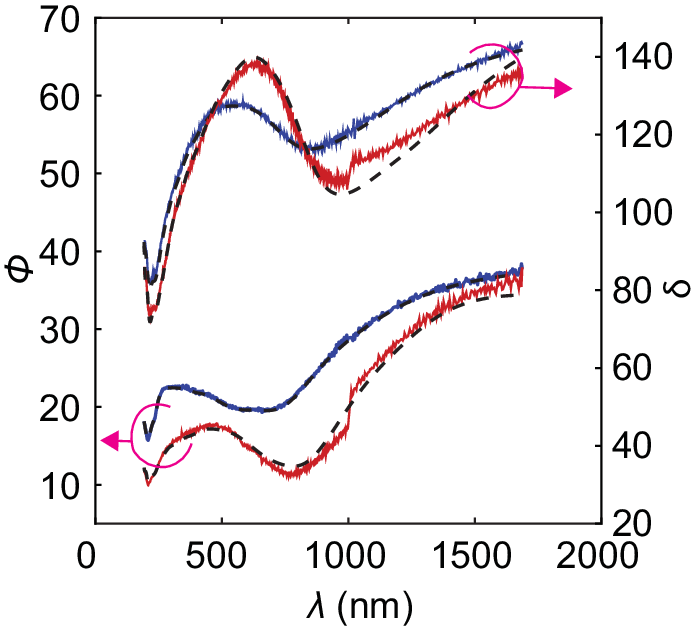}
\caption*{{\red Fig. S6. Spectroscopic ellipsometry measurement for the polished UTe$_2$ sample (blue) and heat-treated UTe$_2$ sample (red). The heating condition is 100 $^\circ$C for 5 minutes in air. The inverse tangential component of the amplitude ratio $\Phi$ and the phase shift $\delta$ are shown at the left and right-axis, respectively. The fits to heat-treated sample using bi-layer models are shown as black dotted lines.} }
\label{figS1}
\end{figure}

\clearpage
\section*{\rom{4}. Magnetic field dependence in S1-A.} \label{Appendix:BField_ZBP_S1A}
\begin{figure}[h]
\includegraphics[width=85 mm]{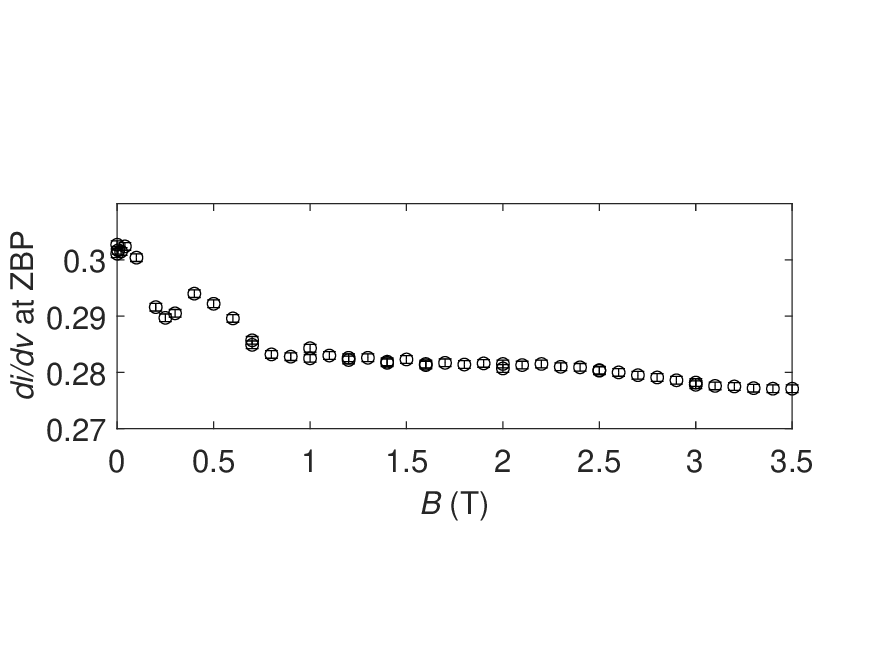}
\caption*{FIG. S7. Magnetic field dependence of the peak value of the zero-bias-peak ($di/dv$ at ZBP) in S1-A.}
\label{ZBP_Bdep}
\end{figure}

\clearpage
\section*{\rom{5}. The upper critical field of samples}
\begin{figure}[h]
\includegraphics[width=85 mm]{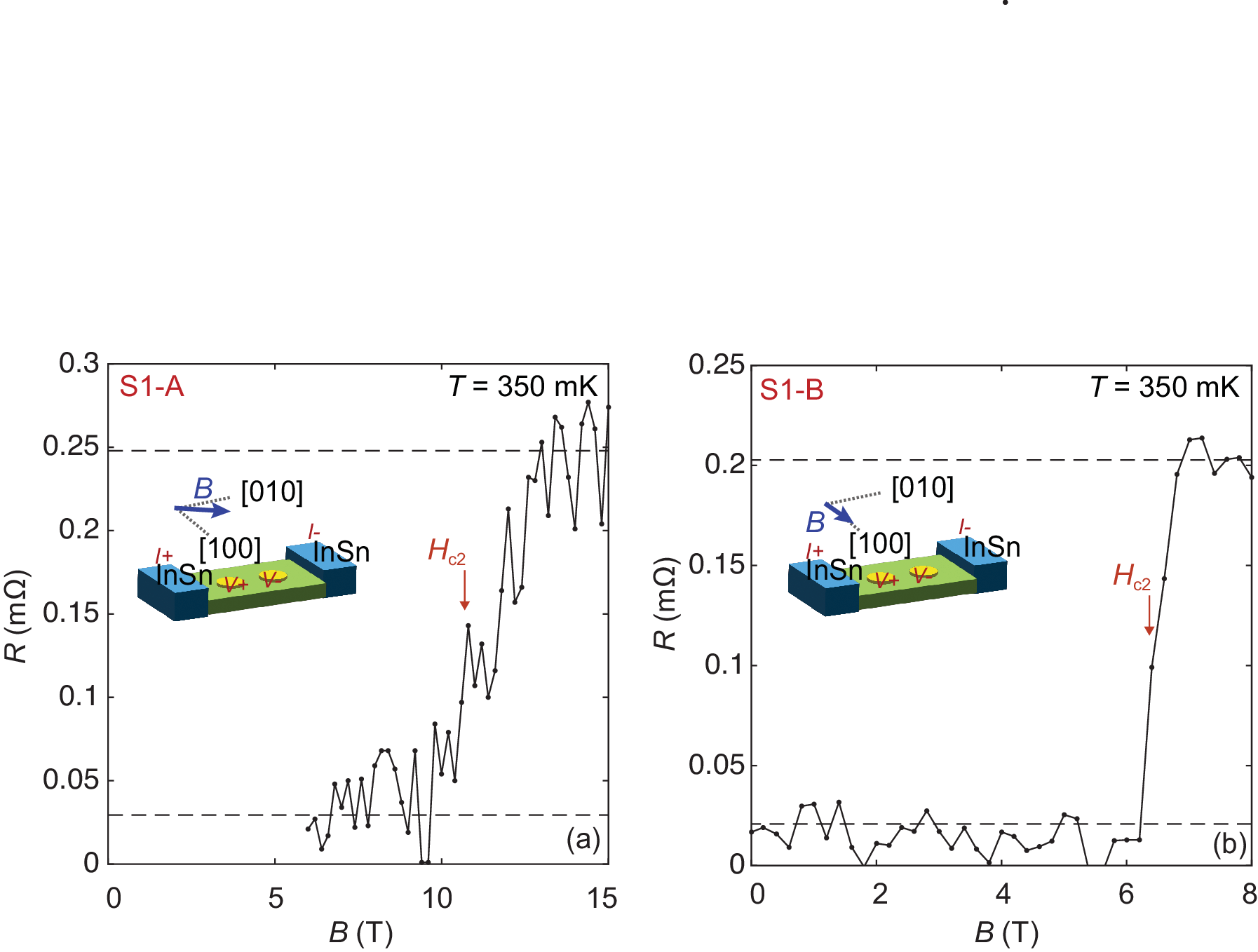}
\caption*{FIG. S8. (a) and (b) shows the resistance as a function of $B$ field for S1-A and S1-B, respectively. For S1-A, $B$ field is applied in-plane, 15 $^\circ$ off from the $b$-axis.   For {\red S1-B}, $B$ field is applied in-plane along $a$-axis. Ohmic contacts are used as current leads and point-contacts are used as voltage leads. {\red The dashed lines show the average values of first and last 7 data points in the curve, respectively, taken as the value for the resistance in superconducting states and normal states.} $H_{c2}$ is defined by the 50\% of transitions.}
\label{Hc2}
\end{figure}

For S2-A and S2-B, $H_{c2}$ is estimated to be $\sim$ 9 T. The $B$ field is applied out-of-plane, which is close to [111].  $H_{c2}$ along [011] and [110] is close to 9 T, and we expect a smooth evolution of $H_{c2}$ between [011] and [110] \cite{Ran2:2019}.
\clearpage

\clearpage
\section*{\rom{6}. S1-A}
\begin{figure}[h!]
\includegraphics[width= 85 mm]{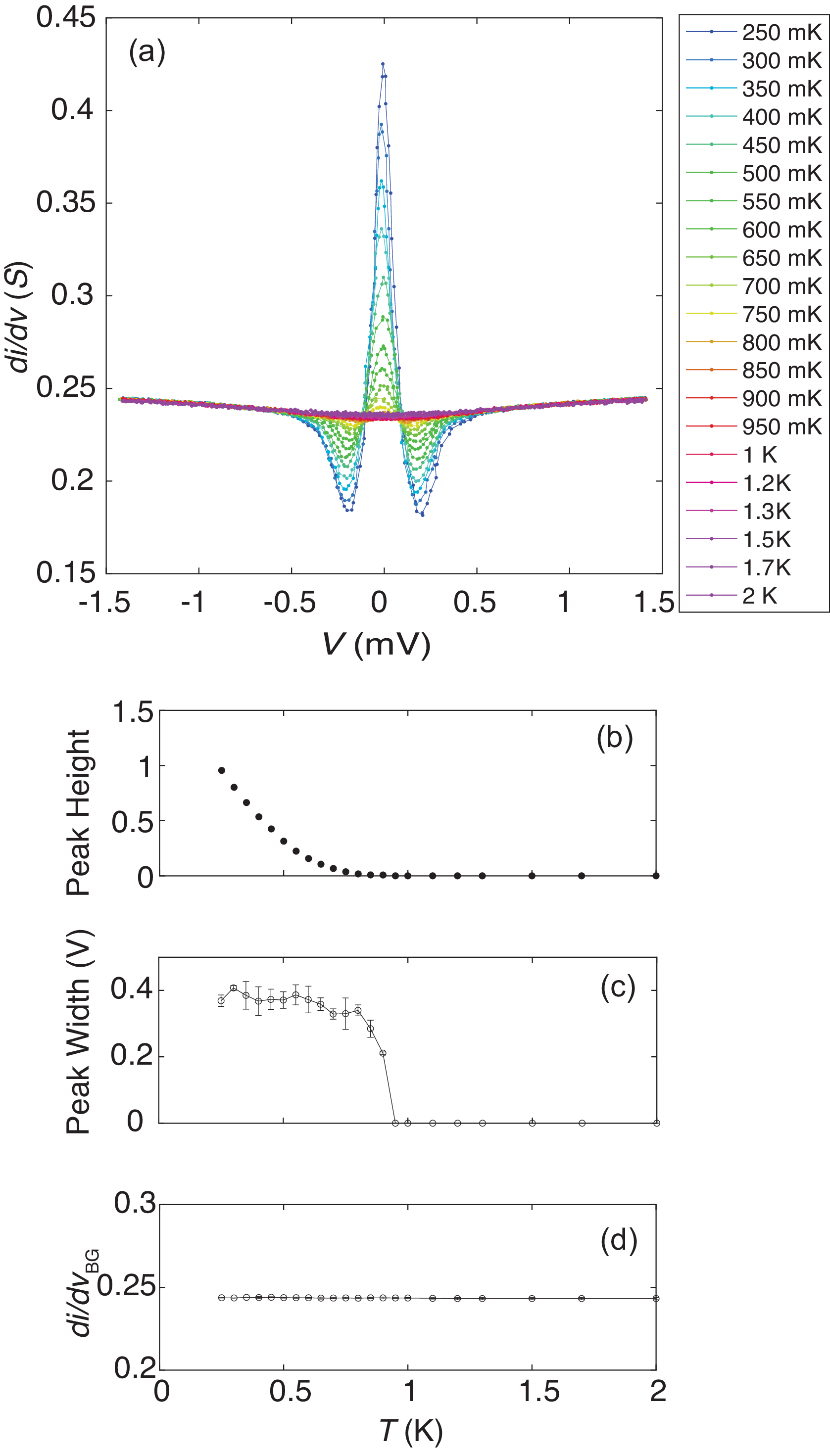}
\caption*{{\red FIG. S9. Temperature dependence of the spectra in S1-A. (a) shows the temperature dependence of the raw data.  (b) shows the extracted peak height as a function of temperature. The peak height here is defined by the maximum value of spectra subtracted by 1 in the normalized $di/dv$ spectra. (c) shows the extracted peak width as a function of temperature. The peak width is defined by the distance between the dips. (d) shows the background at high bias regime. It is obtained by averaging 10 data points from each end.  }}
\label{figS1}
\end{figure}

\newpage
\section*{\rom{7}. Theory} \label{Appendix:Theory}
\begin{spacing}{1.5}

{\red
Within the generalized Blonder-Tinkham-Klapwijk (BTK) theory \cite{BTK:1982,Tanaka:1995} the conductance of a normal metal-superconductor junction is given by
\begin{equation}
    \sigma_{\text{BTK}}(E) = \int \left(1 - \frac{1}{2}\sum_{\sigma,\sigma'}(|b_{\sigma,\sigma'}(\boldsymbol{k_\parallel},E)|^2 - |a_{\sigma,\sigma'}(\boldsymbol{k_\parallel},E)|^2)\right)d\boldsymbol{k_\parallel}\,. \label{eq:BTK}
\end{equation}
where $a_{\sigma,\sigma'}(\boldsymbol{k_\parallel},E)$ and $b_{\sigma,\sigma'}(\boldsymbol{k_\parallel},E)$ give the probability amplitude that an incident spin-$\sigma$ electron with momentum component $\boldsymbol{k_\parallel}$ parallel to the interface and energy $E$ is Andreev reflected as a spin-$\sigma'$ hole or normal reflected as a spin-$\sigma'$ electron.
Our generalized BTK theory employs a number of standard simplifying assumptions~\cite{Daghero_2010}: the Fermi surfaces in both materials are assumed to be spherical with the same radius $k_F$ and isotropic effective mass $m^\ast$; we model the interface by a $\delta$-function potential of strength $\hbar^2k_FZ/2m^\ast$; and we neglect the variation of the gap near the surface. Relaxing these assumptions may alter the quantitative values of our fit parameters but is not expected to qualitatively alter our conclusions; in particular, the topological origin of the ZBP makes this feature immune to details of the system. We performed a least-squares fit to the experimental data, using as our fitting parameters the relative strength of the different components, the overall gap amplitude $\Delta$, the interface barrier strength $Z$, and the broadening parameter $\Gamma$.   

The probability amplitudes in Eq.~\ref{eq:BTK} also appear in the scattering wavefunctions describing the injection of a spin-$\sigma$ electron , 
\begin{equation}
\Psi_{\sigma}(\boldsymbol{k_\parallel},r) = \begin{cases}
\psi^{(N)}_{e,\sigma}e^{ik_\perp r} + \sum_{\sigma'}b_{\sigma,\sigma'}(\boldsymbol{k_\parallel},E)\psi^{(N)}_{e,\sigma'}e^{-ik_\perp r} + \sum_{\sigma'}a_{\sigma,\sigma'}(\boldsymbol{k_\parallel},E)\psi^{(N)}_{h,\sigma'}e^{ik_\perp r} & r<0\\
\sum_{n}c_{\sigma,n}(\boldsymbol{k_\parallel},E)\psi^{(S)}_{e,n}(k_\perp,\boldsymbol{k_\parallel})e^{ik_\perp r} + \sum_{n}d_{\sigma,n}(\boldsymbol{k_\parallel},E)\psi^{(S)}_{h,n}(-k_\perp,\boldsymbol{k_\parallel})e^{-ik_\perp r} & r>0
\end{cases}\label{eq:scattwf}
\end{equation} 
where $\psi^{(N)}_{e(h),\sigma}$ is the spinor corresponding to a spin-$\sigma$ electron (hole) in the normal lead, while $\psi^{(SC)}_{e(h),n=1,2}(\boldsymbol{k})$ describes the two electronlike (holelike) excitations in the superconductor with wavevector $\boldsymbol{k}$ and energy $E$. The coefficients are determined from the boundary conditions
\begin{gather}
\Psi_\sigma(\boldsymbol{k_\parallel},r=0^{-}) = \Psi_\sigma(\boldsymbol{k_\parallel},r=0^{+})\\
\partial_r\Psi_\sigma(\boldsymbol{k_\parallel},r=0^{-}) -\partial_r \Psi_\sigma(\boldsymbol{k_\parallel},r=0^{+}) = k_FZ\Psi_\sigma(\boldsymbol{k_\parallel},r=0^{+})
\end{gather}
The chief difficulty in solving these equations for the coefficients in Eq.~\ref{eq:scattwf} is the nontrivial matrix structure of the triplet gap function $({\bf d}_{\boldsymbol{k}}\cdot\hat{\boldsymbol{\sigma}})i\hat\sigma_y$, where $\hat{\boldsymbol{\sigma}}$ is the vector of Pauli matrices $\hat{\sigma}_{\mu=x,y,z}$, and ${\bf d}_{\bf k}$ is the so-called ${\bf d}$-vector.
For a general triplet pairing state in UTe$_2$, we expect that all elements of the pairing matrix are nonzero. This implies that the electronlike and holelike spinors in general have four nonzero elements, and so the boundary conditions describe a system of eight simultaneous equations, resulting in highly complicated expressions for the conductance.

We can nevertheless make analytic progress by assuming a unitary triplet state, i.e. we restrict ourselves to pairing where ${\bf d}_{\boldsymbol{k}}\times {\bf d}_{\boldsymbol{k}}^\ast = 0$. In the context of UTe$_2$ this implies a time-reversal-symmetric state. Assuming spin-rotation invariance in the normal metal, we are then able to choose a spin-quantization axis \emph{for each individual $\boldsymbol{k_\parallel}$} such that the ${\bf d}$-vectors at $\boldsymbol{k}_o=(k_\perp,\boldsymbol{k_\parallel})$ and $\boldsymbol{k}_i=(-k_\perp,\boldsymbol{k_\parallel})$ (i.e. the ``outgoing'' and ``incoming'' wavevectors in the wavefunction Ansatz Eq.~\ref{eq:scattwf}) lie in the $x$-$y$ plane, and hence involve only so-called equal-spin pairing. This reduces the $4\times4$ Bogoliubov-de Gennes equation to two $2\times 2$ matrices, one for each spin orientation.

Denoting the ${\bf d}$-vectors at $\boldsymbol{k}_o$ and $\boldsymbol{k}_i$ in the original basis by ${\bf d}_{o}$ and ${\bf d}_{i}$, respectively, we have in the rotated basis
\begin{align}
  {\bf d}_o^\prime & = \frac{d_{i,z}({\bf d}_{o}^2 - d_{o,z}^2) - d_{o,z}(({\bf d}_{i}\cdot{\bf d}_{o} - d_{i,z}d_{o,z}))}{|\hat{\boldsymbol{z}}\times({\bf d}_{o}\times{\bf d}_{i})|}\hat{\bf x} - d_{o,z}\frac{|{\bf d}_{o}\times{\bf d}_{i}|}{|\hat{\boldsymbol{z}}\times({\bf d}_{o}\times{\bf d}_{i})|}\hat{\bf y}\\
  {\bf d}_i^\prime & = \frac{-d_{o,z}({\bf d}_{i}^2 - d_{i,z}^2) + d_{i,z}(({\bf d}_{i}\cdot{\bf d}_{o} - d_{i,z}d_{o,z}))}{|\hat{\boldsymbol{z}}\times({\bf d}_{o}\times{\bf d}_{i})|}\hat{\bf x} - d_{i,z}\frac{|{\bf d}_{o}\times{\bf d}_{i}|}{|\hat{\boldsymbol{z}}\times({\bf d}_{o}\times{\bf d}_{i})|}\hat{\bf y}
\end{align}
The pairing potential at e.g. $\boldsymbol{k}_o$ is therefore transformed as
\begin{equation}
  {\bf d}_o\cdot\hat{\boldsymbol{\sigma}}i\hat{\sigma}_y  \rightarrow {\bf d}^\prime_o\cdot\hat{\boldsymbol{\sigma}}i\hat{\sigma}_y = \begin{pmatrix}
    -d^\prime_{o,x} + id^\prime_{o,y} & 0 \\ 0 & d^\prime_{o,x} + id^\prime_{o,y}
  \end{pmatrix} = \begin{pmatrix}
    -|{\bf d}_o|e^{-i\varphi_o} & 0 \\
    0 & |{\bf d}_o|e^{i\varphi_o}
    \end{pmatrix} \label{eq:diagonal}
\end{equation}
which allows us to define a phase $\varphi$ of the pairing potential.

In terms of the rotated ${\bf d}$-vectors, we can now obtain explicit expressions for the Andreev and normal reflection coefficients for the injection of a spin-$s$ electron (with quantization axis  ${\bf d}_{o}\times{\bf d}_{i}$) as
\begin{align}
  a_{ss}(\boldsymbol{k_\parallel},E) & = \frac{\exp(-i\phi_{o,s})u_iv_o}{(1+\tilde{Z}^2)u_ou_i - \tilde{Z}^2v_ov_i\exp(-is[\phi_{i}-\phi_{o}])}\\
    b_{ss}(\boldsymbol{k_\parallel},E) & = \frac{\tilde{Z}(i+\tilde{Z})[v_ov_i\exp(i\phi_{i,s}-i\phi_{o,s}) - u_ou_i]}{(1+\tilde{Z}^2)u_ou_i - \tilde{Z}^2v_ov_i\exp(-is[\phi_{i}-\phi_{o}]}
\end{align}
where $\tilde{Z} = Zk_F/k_\perp$, 
\begin{gather}
  u_{o/i} = \sqrt{\frac{E+\Omega_{o/i}}{2E}}\,, \qquad   v_{o/i} = \sqrt{\frac{E-\Omega_{o/i}}{2E}}\,, \quad \Omega_{o/i} = \sqrt{E^2-|{\bf d}_{o/i}|^2}\,,
\end{gather}
and we can define the phases as in Eq.~\ref{eq:diagonal}.
These formulas for the reflection coefficients are essentially identical to those for an anisotropic spin-singlet superconductor~\cite{KashiwayaTanaka_2000}.

\subsubsection*{Surface bound states}

We can obtain the surface bound states at the bare surface (i.e. no lead) by making the analytic continuation $E\rightarrow E + i0^{+}$ and searching for poles of the scattering amplitudes in the limit $Z\rightarrow\infty$. For $|E|<\min\{|{\bf d}_i|,|{\bf d}_o|\}$, we find that the surface bound states are given by the solution of the equation
\begin{equation}
|{\bf d}_i||{\bf d}_o| - (E-i\sqrt{|{\bf d}_o|^2 - E^2})(E-i\sqrt{|{\bf d}_i|^2 - E^2})\exp(is[\phi_{i}-\phi_{o}]) = 0 \label{eq:surfacebs}
\end{equation}
In particular, a phase difference of $\pi$ implies that $E=0$ is a solution (i.e. a zero-energy state), whereas for a phase difference of $0$ no solution is possible. In the special case where $|{\bf d}_i|=|{\bf d}_o|=\Delta$, the bound state energies are given by
\begin{equation}
E_{\pm} = \pm \Delta\cos((\phi_{i}-\phi_{o})/2)
\end{equation}

In the main text, it was claimed that the surface bound states for a gap with a dominant $p_y$-wave component are very similar to those obtained for a purely $p_y$-wave gap. To illustrate this, let us consider the surface bound states of an $A_{u}$ state with ${\bf d}_{\boldsymbol{k}} = ak_x\hat{\bf x} + bk_y\hat{\bf y} + ck_z\hat{\bf z}$ at the (001) and (010) surfaces. At these surfaces, the phase differences are given by
\begin{eqnarray}
  \text{$(010)$ surface:} &\qquad & \phi_{i} - \phi_o = \pi - 2\arctan\left(\frac{\sqrt{(ak_x)^2+(ck_z)^2}}{|b|\sqrt{k_F^2 - k_x^2 - k_z^2}}\right)\\
  \text{$(001)$ surface:} &\qquad & \phi_{i} - \phi_o = - 2\arctan\left(\frac{|c|\sqrt{k_F^2 - k_x^2 - k_y^2}}{\sqrt{(ak_x)^2 + (bk_y)^2}}\right)    
  \end{eqnarray}
where we have expressed this in terms of the momentum components parallel to the interface. We see that for $|b|\gg |a|, |c|$, the phase shift at the $(010)$ surface is close to $\pi$ except very close to the edge of the projected Fermi surface; on the other hand, the phase shift for the $(001)$ surface is close to zero except near the zone centre or along the line $k_y=0$. This supports our argument that the phase shifts for a pairing state with dominant $p_y$-wave component are very similar to those for a purely $p_y$-wave pairing state. 

To further illustrate this, we plot the bound state spectrum at the $(010)$ and $(001)$ surfaces in Figs. S10 and S11, respectively, comparing the case of purely $p_y$-wave pairing, dominant $p_y$-wave pairing, and isotropic pairing. As can be seen, the dominant $p_y$-wave pairing state has aspects of both the purely $p_y$-wave pairing and the isotropic pairing: surface states for all momenta within the projected Fermi surface, and a Dirac cone in the zone centre. However, the edge states of the dominant $p_y$-wave state are also quite similar to that for the purely $p_y$-wave gap: close to zero energy for the $(010)$ surface, and also close to the gap edge for the $(001)$ surface.
For sufficiently high temperature and broadening parameter, the differences between the surface spectra for the dominant and purely $p_y$-wave states cannot be resolved in the tunneling conductance, and the purely $p_y$-wave state gives a qualitatively correct guide to the tunneling conductance.
}
\begin{figure}
  (a)\includegraphics[width=0.3\textwidth]{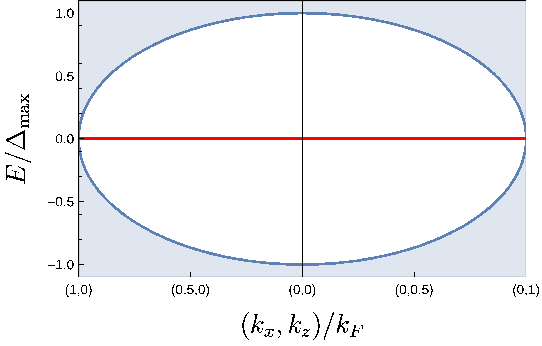}(b)\includegraphics[width=0.3\textwidth]{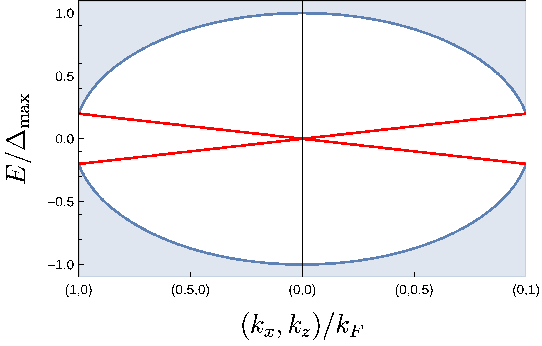}(c)\includegraphics[width=0.3\textwidth]{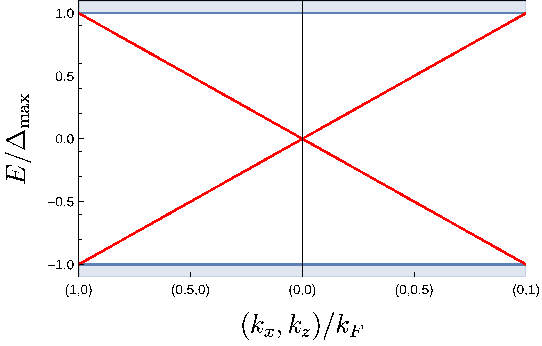}
  \caption*{FIG. S10. Bound states at the (010) surface for (a) purely $p_y$-wave pairing, (b) an anisotropic $A_u$ state with $b=5a=5c$, (c) an isotropic $A_u$ state with $b=a=c$. The continuum region is shaded, while the surface bound states are shown as the red lines.}
  \label{fig:surfacestates001}
\end{figure}

\begin{figure}
  (a)\includegraphics[width=0.3\textwidth]{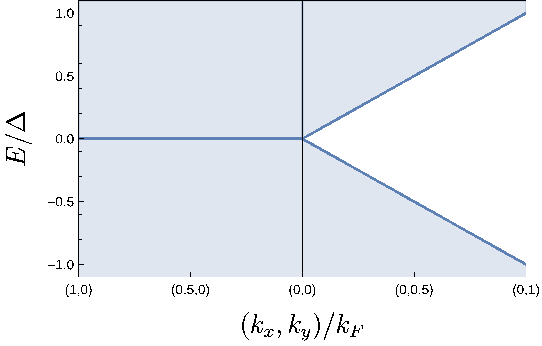}(b)\includegraphics[width=0.3\textwidth]{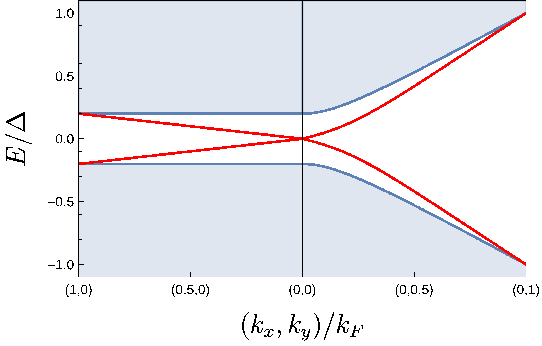}(c)\includegraphics[width=0.3\textwidth]{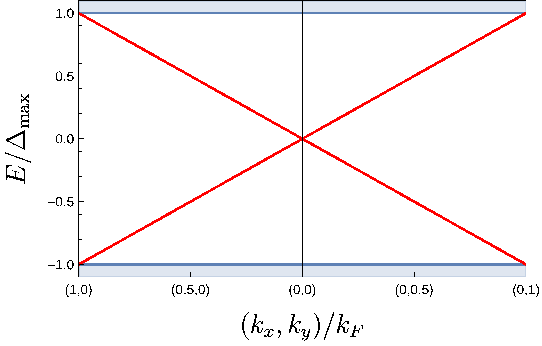}
  \caption*{FIG. S11. Bound states at the (001) surface for (a) purely $p_y$-wave pairing, (b) an anisotropic $A_u$ state with $b=5a=5c$, (c) an isotropic $A_u$ state with $b=a=c$. The continuum region is shaded, while the surface bound states are shown as the red lines.}
  \label{fig:surfacestates001}
\end{figure}

\end{spacing}

\section*{\rom{8}. BTK fits to S1-A and S1-B}

\begin{figure}[h!]
\includegraphics[width= 85 mm]{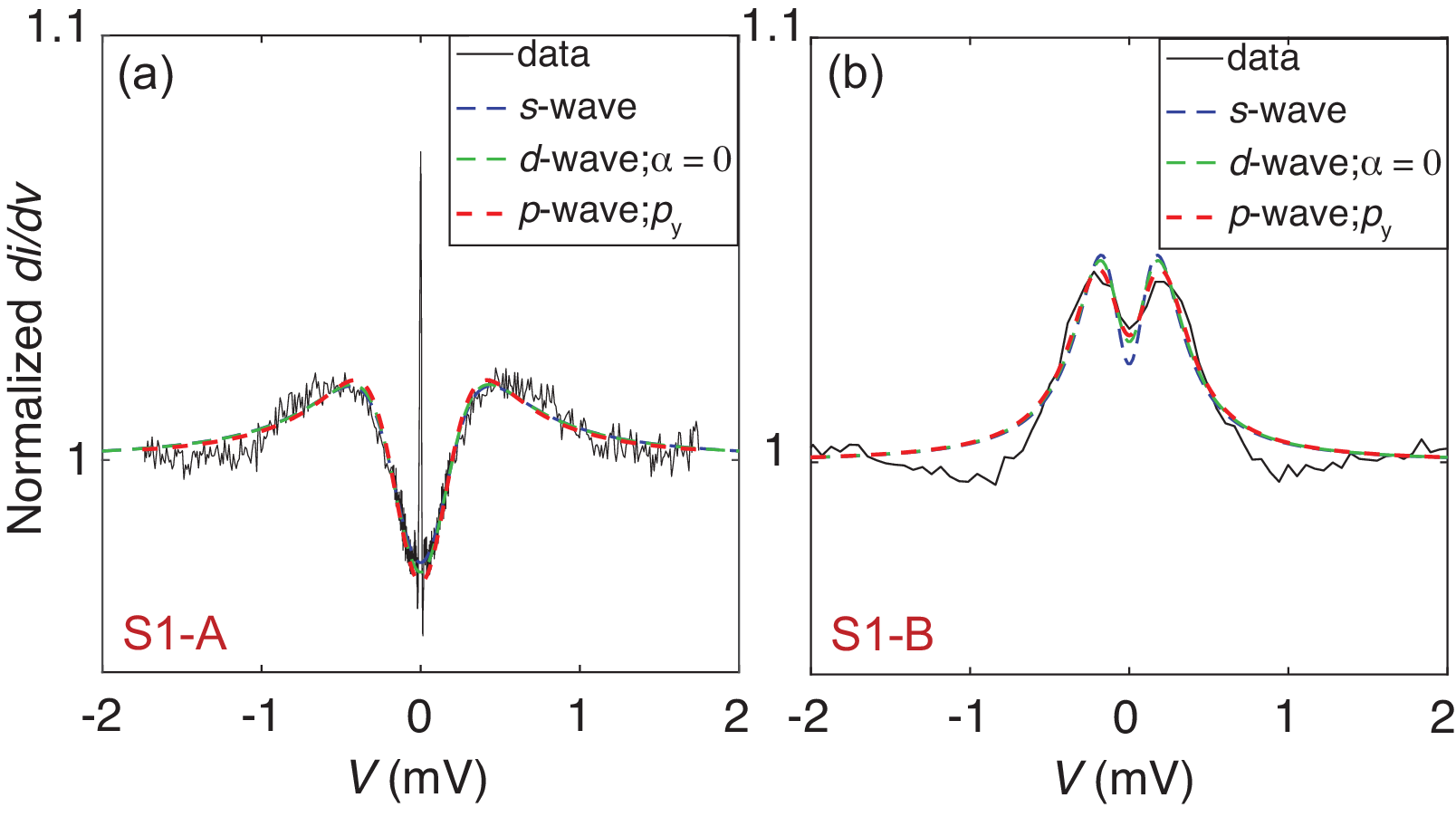}
    \caption*{{\red FIG. S12. BTK fits for S1-A and S1-B using $s$-wave, $d$-wave, and $p$-wave symmetry. To capture the gap-like feature in S1-A and S1-B, $\alpha$ is set to be zero in $d$-wave model, where $\Delta = \cos(2(\theta+\alpha))$. $p_y$ symmetry is used for $p$-wave model.   The fitting parameters $\Delta$, $\Gamma$, and $Z$ are shown in Table S2. }}
\label{figS1}
\end{figure}

\end{spacing}
\begin{table}[h!]
 \centering
 
  \begin{tabular} {  |p{2cm} | p{2cm} | p{3cm} | p{3cm} | p{3cm}|}
   \hline
    Sample & Symmetry & $\Delta$ (meV) & $\Gamma$ (meV)  & $Z$ \\ \hline \hline
    \multirow{3}{*}{S1$\mbox{-}$A} &$s$-wave &$0.22\pm0.052$  & $0.29\pm0.084$ & $0.91\pm0.027$ \\ \cline{2-5}
    &$d$-wave & $0.15\pm0.005$ &$0.30\pm0.005$ &$0.84\pm0.013$ \\ \cline{2-5}
    &$p$-wave &$0.32\pm0.007$ &$0.24\pm0.005$ &$0.63\pm0.008$ \\ \hline
 \multirow{3}{*}{S1$\mbox{-}$B} &$s$-wave &$0.12\pm0.008$  & $0.14\pm0.006$ & $0.50\pm0.001$ \\ \cline{2-5}
    &$d$-wave & $0.18\pm0.012$ &$0.14\pm0.012$ &$0.50\pm0.001$ \\ \cline{2-5}
    &$p$-wave &$0.26\pm0.014$ &$0.13\pm0.015$ &$0.32\pm0.02$ \\ \hline

\end{tabular}

\caption*{\label{Table2} \red{Table S2. The fitting parameters $\Delta$, $\Gamma$, and $Z$ for BTK fits in Fig. S10, using $s$-wave, $d$-wave, and $p$-wave symmetry. }} 

\end{table}

\end{document}